\documentclass[3p,sort&compress,twocolumn]{elsarticle}

\usepackage{amsfonts,amssymb,amsmath,times}%
\usepackage{graphicx}
\usepackage{bm}
\usepackage{enumerate}
\usepackage{color}

\journal{Frontiers of Physics}
\begin{document}

\title{Cross and joint ordinal partition transition networks for multivariate time series analysis}

\author[SH]{Heng Guo}
	\address[SH]{Department of Physics, East China Normal University, Shanghai,
200062, China}

\author[SH]{Jiayang Zhang}

\author[SH]{Yong Zou\corref{corrauthor}} 
\cortext[corrauthor]{Corresponding author}
\ead{yzou@phy.ecnu.edu.cn}

\author[SH]{Shuguang Guan\corref{corrauthor}} 
\cortext[corrauthor]{Corresponding author}
\ead{sgguan@phy.ecnu.edu.cn}

\date{\today}

\begin{abstract}
We propose to construct cross and joint ordinal pattern transition networks from multivariate time series for two coupled systems, where synchronizations are often present. In particular, we focus on phase synchronization, which is one of prototypical scenarios in dynamical systems. We systematically show that cross and joint ordinal patterns transition networks are sensitive to phase synchronization. Furthermore, we find that some particular missing ordinal patterns play crucial roles in forming the detailed structures in the parameter space whereas the calculations of permutation entropy measures often do not. We conclude that cross and joint ordinal partition transition network approaches provide complementary insights to the traditional symbolic analysis of synchronization transitions.
\end{abstract}

\maketitle

\section{Introduction}
Complex network theory provides an important paradigm for understanding the structural properties of complex systems composed of different interacting entities. During the last decade, various complex network approaches have been proposed to extract useful insights from time series data \cite{Donner2011IJBC}. Depending on slightly different definitions of nodes and links of network representations for time series, there are methods based on recurrences \cite{Donner2010NJP,Marwan2009}, visibility conditions \cite{Lacasa2008}, cycles detections \cite{zhangPRL2006} and correlation networks \cite{yangPhA2008}. Some successful applications of these methods include time series from climate, sunspots, oil-water flows, and financial market \cite{Donner2011IJBC,Donges2011PNAS,Zou2012bChaos,gaoChaos2017,Elsner2009,zouNJP2014,zouNPG2014,ZhangCNSNS2017,CzechowskiChaos2016,ZhangFOP2017a,JiangFOP2017}. The consideration of a time series as a complex network allows a reinterpretation of many network theoretic measures in terms of characteristic phase space properties of a dynamical system \cite{DonnerEPJB2011}. 

Recently, a growing number of works are focused on transforming time series into networks by ordinal partitions of time series \cite{MichaelChaos2015,KulpChaos2016,KulpChaos2016b,McCulloughChaos2016,SakellariouChaos2016}. The basic idea of ordinal partition network method can be traced back to identifying ordinal patterns of time series \cite{BandtPRL2002,ParlitzSPJST2013,AmigoPTRSA2014}. Given a one-dimensional time series $\{ x(t)\}_{t=1, \cdots, L}$ comprising of $L$ points from a dynamical system, we first reconstruct its phase space by time delay embedding technique which yields $\vec{x}(t) = [x(t), x(t+\tau), \cdots, x(t+(D_x-1)\tau)]$, where $D_x$ and $\tau$ are embedding dimension and delay respectively \cite{Takens1981,Kantz97}. The next step is to compute the rank order of each embedded vector $[x(t), x(t+\tau), \cdots, x(t+(D_x-1)\tau)]$, which is conveniently denoted by a symbol $\pi_x(t)$. When sliding windows from $t=1$ to $N = L - (D_x - 1)\tau$ in the embedded space, a symbolic representation of the trajectory $\pi_x(t)$ is produced. Following the symbolic representation, one traditional approach is to compute permutation entropy $\mathcal{H}_O$ based on the frequency plot of order patterns. Generally speaking, for a time series generated by a stochastic process for $N \to \infty$, it is known that all $D_x!$ patterns almost occur with equal probabilities, which yields the maximal value of $\mathcal{H}_O$. However, for a time series produced by deterministic dynamics, the frequency plot of the $D_x!$ patterns is not uniform, which leads to reduced value of $\mathcal{H}_O$.  In some cases, a set of patterns may never occur and these missing patterns are often called forbidden patterns, which provide important information for quantifying determinism in time series data \cite{AmigoEPL2007,AmigoEPL2008,RossoEPJB2012,RossoPhyA2012,KulpChaos2014}. Then, the level of determinism of a time series from a particular dynamical system may be suggested by permutation entropy $\mathcal{H}_O$,  which consists in very well established statistical measures in nonlinear time series analysis \cite{BandtPRL2002,AmigoPTRSA2014,politiPRL2017}. Some applications include characterizations of the difference between healthy and patients from EEG data \cite{MichaelChaos2015,KulpChaos2016}. It is worth noting that complications may arise in real time analysis because missing ordinal patterns might be related to finite time length during the period of observation and correlated stochastic processes, which require some revised methods for the detection of determinism in relatively short noisy data \cite{AmigoEPL2007,AmigoEPL2008,RossoEPJB2012,RossoPhyA2012,KulpChaos2014}. 

The primary idea of ordinal partition transition network takes into account the inhomogeneous evolutionary behavior among the ordinal patterns \cite{MichaelChaos2015,KulpChaos2016}, which provides complementary information on the standard ordinal symbolic analysis of time series. Most of these works have focused on univariate time series $\{x(t)\}$. Embedding parameters $D_x$ and $\tau$ have crucial impacts on the resulting ordinal partition transition networks, especially for forbidden patterns \cite{KulpChaos2016b,McCulloughChaos2016,SakellariouChaos2016}. In addition, multivariate time series are ubiquitous in nature, ranging from stock markets and climate sciences. In a recent work, we proposed to construct ordinal partition transition networks from multivariate data \cite{zhangSciRep2017}. The resulting network is a directed and weighted network characterizing the pattern transition properties of time series in its associated velocity space. This novel approach has been successfully applied to capture phase coherence to non-phase coherence transitions and to characterize paths to phase synchronization, showing complementary insight to the traditional symbolic analysis of nonlinear time series analysis. 

In this work, we further extend these ideas \cite{zhangSciRep2017} to construct cross and joint ordinal partition transition networks for two coupled systems. We show that both cross and joint ordinal pattern transition networks are able to capture synchronization transitions, in particular, focusing on the transitions to phase synchronization (PS) which is one of paradigmatic types in synchronization phenomena \cite{Pikovsky_Kurths_synchr,Osipov2003,ZhangFOP2017,ChenFOP2017,HuangFOP2016,YingFOP2017}.  The outline of this paper is as follows: first, we illustrate cross and joint network construction approaches and then introduce two entropy measures to quantify the inhomogeneous frequencies of ordinal patterns and their transitions in Sec. \ref{secCOPTJ}. We apply these two entropy measures to characterize the synchronization transitions with both unidirectional and bidirectional coupling schemes in Sec. \ref{secSync}, and then some conclusions are drawn in Sec. \ref{secCons}. 

\section{Cross and joint ordinal partition transition networks} \label{secCOPTJ}
\subsection{Network constructions} \label{secRungeK}
The ordinal partition transition networks are illustrated by the following coupled R\"ossler systems~\cite{zhengPRE2000,mamenPLA2004}, which shows transition scenarios to phase synchronization for various coupling schemes. In particular, PS has been observed for both unidirectional and bidirectional couplings. Note that the network construction method is not restricted by the following model. The ordinary differential equations (ODEs) of the coupled system read
\begin{align}\label{eq_roessler}
\begin{cases}
\dot{x}_1 &= -(1.0 - \Delta) y_1 - z_1 + \kappa_1 (x_2 - x_1), \\
\dot{y}_1 &= (1.0 - \Delta)x_1 + 0.15 y_1, \\
\dot{z}_1 &= 0.2 + z_1 (x_1 - 10.0),
\end{cases}\\
\begin{cases}
\dot{x}_2 &= -(1.0 + \Delta) y_2 - z_2 + \kappa_2 (x_1 - x_2), \\
\dot{y}_2 &= (1.0 + \Delta) x_2 + 0.15 y_2, \\
\dot{z}_2 &= 0.2 + z_2 (x_2 - 10.0),
\end{cases}
\end{align}
where $\omega_{1} = 1.0 - \Delta, \omega_2 = 1.0 + \Delta$ are frequencies for the systems, and $\kappa_{1,2}$ are coupling strength. In this model, we distinguish a unidirectional coupling case from a bidirectional coupling one because the threshold values to PS have been reported in the literature \cite{zhengPRE2000,mamenPLA2004}. More specifically the unidirectional coupling can be achieved by $\kappa_1 = 0$, namely, a drive-response scheme \cite{zhengPRE2000} and bidirectional coupling is often chosen as $\kappa_1 = \kappa_2 = \kappa$ \cite{mamenPLA2004}. We numerically integrate the ODEs using the fourth-order Runge-Kutta method with random initial conditions, and the integration step $h = 0.01$. The first $10000$ transient data points are discarded and time series consisting of $N=800000$ data points are analyzed.

\begin{figure}
	\centering
	\includegraphics[width=\columnwidth]{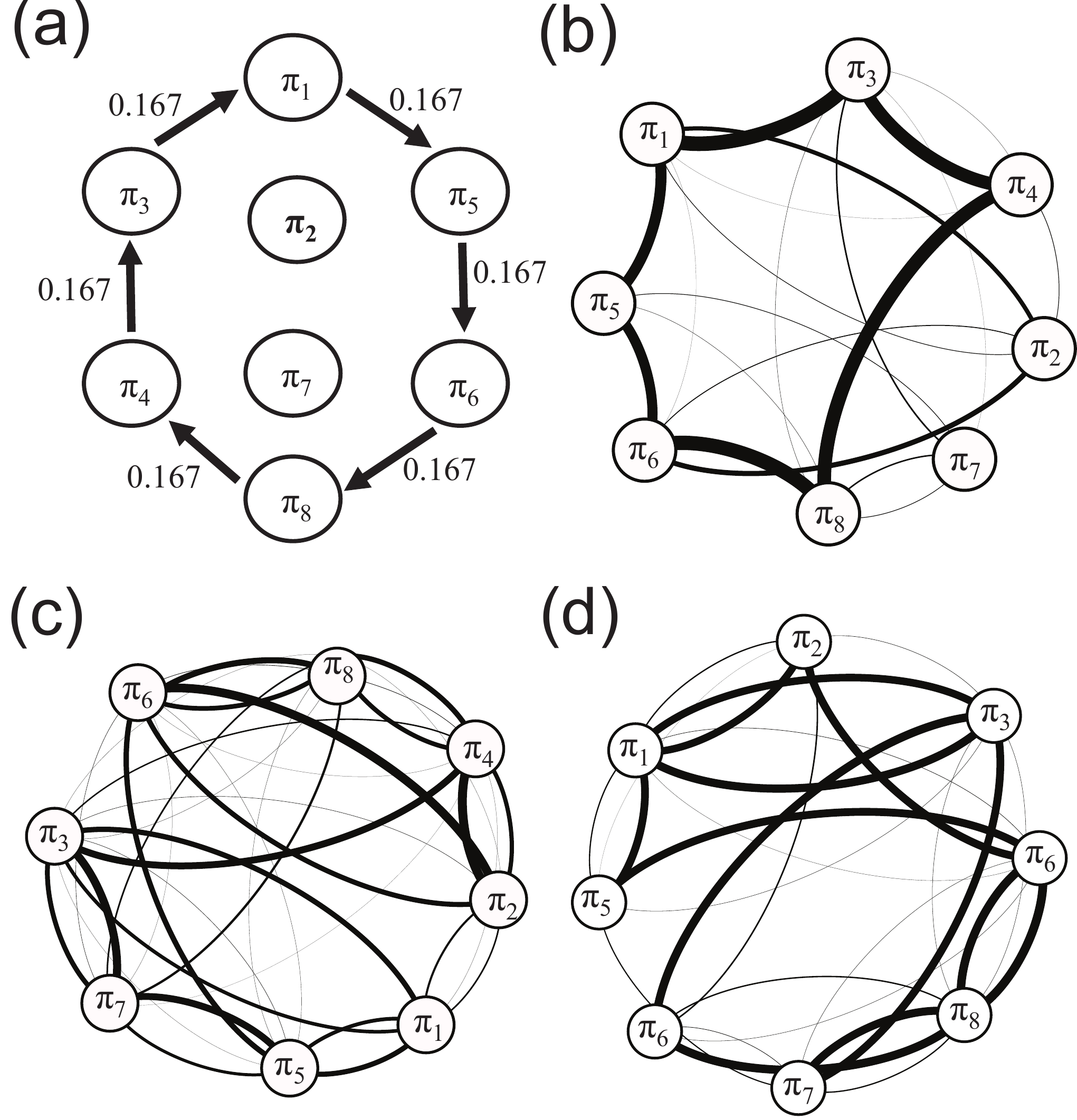}
	\caption{\small{Ordinal pattern transition network, where the thickness (and values) on network links represent the corresponding transition frequency between two patterns. (a) A single chaotic R\"ossler system. (b) Cross ordinal pattern transition network (COPT) for two coupled R\"ossler systems, (c) an alternative version of COPT, and (d) joint ordinal pattern transition network (JOPT). For (b)-(d), the directions of links have been suppressed for better visualizations. In addition, the coupling strength $\kappa_{1,2}$ is in the non-sync regime, namely, $\kappa_{1} = 0, \kappa_2 = 0.01$. } \label{fig:rosCOPT}}
\end{figure}

{\emph{Ordinal Pattern Transition network. }} We start the idea by constructing the ordinal pattern transition network for a single system, using the first system (Eqs. \eqref{eq_roessler}, $\omega_1 = 1.0, \Delta = 0, \kappa_{1,2} = 0$) as an example \cite{zhangSciRep2017}. Given time series $(x(t), y(t), z(t))$ in the corresponding three dimensional phase space ($n = 3$), the ordinal pattern transition network is reconstructed based on the signs of the increments of each each variablee $(\Delta x(t), \Delta y(t), \Delta z(t))$, where $\Delta x(t) = x(t+1) - x(t)$, $\Delta y(t) = y(t+1) - y(t)$, and $\Delta z(t) = z(t+1) - z(t)$. In particular, the definitions of the ordinal patterns based on the increment series capture the variations of the trajectory in its associated velocity space. The definition of patterns $\Pi(t) \in (\pi_1, \cdots, \pi_i), i = 1, \cdots, 8$ are enumerated in Tab. \ref{tab:3D}. As time evolves, the transition behavior between patterns are illustrated in Fig. \ref{fig:rosCOPT}(a). The deterministic transitions are explained by the ordinal partitions of phase space by null-clines. A transition between two patterns means the trajectory crosses a null-cline, which leads to a local maximum or minimum. Because of the continuity of the system in phase space, we only observe the transition route $\pi_1 \to \pi_5 \to \pi_6 \to \pi_8 \to \pi_4 \to \pi_3 \to \pi_1$, yielding two missing patterns $\pi_2$ and $\pi_7$ as shown in Fig. \ref{fig:rosCOPT}(a). These details  have been well illustrated in \cite{zhangSciRep2017}. 
\begin{table}[htb]
\centering
\begin{tabular}{|c|c|c|c|c|c|c|c|c|}
\hline
$\Pi$      & $\pi_1$ & $\pi_2$ & $\pi_3$ & $\pi_4$ & $\pi_5$ & $\pi_6$ & $\pi_7$
& $\pi_8$\\
\hline
$\Delta x$ & $ + $ & $ + $ & $  +$ & $ +$ & $  - $ & $ - $ & $ -$ & $ -$\\
\hline
$\Delta y$ & $ + $ & $ + $ & $-$ & $ -$ & $ + $ & $ + $ & $ -$ & $ -$\\
\hline
$\Delta z$ & $ + $ & $ - $ & $ +$ & $ -$ & $ + $ & $ - $ & $ +$ & $ -$\\
\hline
\end{tabular}
\caption{Definitions of ordinal patterns of the three dimensional time series $(x(t), y(t), z(t))$, where $\Delta x = x(t+1) - x(t)$, $\Delta y = y(t+1) - y(t)$, and $\Delta z = z(t+1) - z(t)$~\cite{zhangSciRep2017}.
\label{tab:3D}}
\end{table}

Generalizing the above ideas from a single system to two coupled systems $(\kappa_{1,2} \neq 0)$, we propose two different ways to construct ordinal pattern transition networks. Suppose we have time series $(x_{1}(t), y_{1}(t), z_{1}(t))$ of one system and $(x_{2}(t), y_{2}(t), z_{2}(t))$ of the other, we first compute the respective increments of the two systems as $(\Delta x_{1}(t), \Delta y_{1}(t), \Delta z_{1}(t))$ and $(\Delta x_{2}(t), \Delta y_{2}(t), \Delta z_{2}(t))$. Note that the increment series capture the dynamic properties of time series in the difference space and the signs of the each variable reflects either the increasing ($+$) or the decreasing ($-$) trend. Next, we compare the two systems in the following:

{\emph{ Cross ordinal pattern transition network (COPT). }}  A COPT compares the relative speeds between two systems by the signs of $(\Delta x_1(t) - \Delta x_2(t)), (\Delta y_1(t) - \Delta y_2(t))$ and $(\Delta z_1(t) - \Delta z_2(t))$. The pattern definitions of a COPT are shown in Tab. \ref{tab:3DCOPT}. An example of COPT is shown in Fig. \ref{fig:rosCOPT}(b). In this work, we only consider the signs of the same variable from two coupled systems. A further generalization of the pattern definitions of a COPT is to consider the signs of cross-variables, for instance, $(\Delta x_1(t) - \Delta y_2(t)) $ and $(\Delta x_1(t) - \Delta z_2(t))$, respectively. 
\begin{table}[htb]
\centering
\begin{tabular}{|c|c|c|c|c|c|c|c|c|}
\hline
$\Pi$      & $\pi_1$ & $\pi_2$ & $\pi_3$ & $\pi_4$ & $\pi_5$ & $\pi_6$ & $\pi_7$
& $\pi_8$\\
\hline
$\Delta x_1 - \Delta x_2$ & $+ $ & $+ $ & $+$ & $+$ & $ - $ & $ - $ & $-$ & $ - $\\
\hline
$\Delta y_1 - \Delta y_2$ & $ + $ & $ + $ & $ -$ & $ -$ & $ + $ & $ + $ & $ -$ & $ -$\\
\hline
$\Delta z_1 - \Delta z_2$ & $ + $ & $ - $ & $ +$ & $ -$ & $ + $ & $ - $ & $+$ & $ -$\\
\hline
\end{tabular}
\caption{Pattern definitions of a COPT. Note that $``+"$ means a positive value while $``-"$ is for a negative value.  \label{tab:3DCOPT}}
\end{table}
Considering the effects of the different magnitudes of the three variables, we also compute an {\emph{alternative}} COPT by replacing $\Delta x_1(t) - \Delta x_2(t)$ by $\Delta x_1(t) / x_1(t) - \Delta x_2(t) / x_2(t)$, respectively, $\Delta y_1(t) - \Delta y_2(t)$ by $\Delta y_1(t) / y_1(t) - \Delta y_2(t) / y_2(t)$, and $\Delta z_1(t) - \Delta z_2(t)$ by $\Delta z_1(t) / z_1(t) - \Delta z_2(t) / z_2(t)$. An example of the alternative COPT is shown in Fig. \ref{fig:rosCOPT}(c). Comparing Fig. \ref{fig:rosCOPT}(b) to \ref{fig:rosCOPT}(c), the alternative COPT reflects better the non-coherent transitions between ordinal patterns since the coupling strength is in the non-synchronization regime ($\kappa_{1} = 0$ and $\kappa_{2} = 0.01$). In the following, we compute the alternative COPTs without distinguishing these two slightly different versions. 

{\emph {Joint ordinal pattern transition networks (JOPT). }} A JOPT compares the relative speeds between two systems by the signs of $\Delta x_1(t) \cdot \Delta x_2(t), \Delta y_1(t) \cdot \Delta y_2(t)$ and $\Delta z_1(t) \cdot \Delta z_2(t)$ and the pattern definitions of a JOPT are summarized in Tab. \ref{tab:3DJOPT}. An example of JOPT is shown in Fig. \ref{fig:rosCOPT}(d).
\begin{table}[htb]
\centering
\begin{tabular}{|c|c|c|c|c|c|c|c|c|}
\hline
$\Pi$      & $\pi_1$ & $\pi_2$ & $\pi_3$ & $\pi_4$ & $\pi_5$ & $\pi_6$ & $\pi_7$
& $\pi_8$\\
\hline
$\Delta x_1 \cdot \Delta x_2$ & $+ $ & $+ $ & $+$ & $+$ & $ - $ & $ - $ & $-$ & $ - $\\
\hline
$\Delta y_1 \cdot \Delta y_2$ & $ + $ & $ + $ & $ -$ & $ -$ & $ + $ & $ + $ & $ -$ & $ -$\\
\hline
$\Delta z_1 \cdot \Delta z_2$ & $ + $ & $ - $ & $ +$ & $ -$ & $ + $ & $ - $ & $+$ & $ -$\\
\hline
\end{tabular}
\caption{Pattern definitions of a JOPT. Note that $``+"$ means a positive value while $``-"$ is for a negative value.  \label{tab:3DJOPT}}
\end{table}
In contrast to cross ordinal patterns, we notice that the joint ordinal patterns represent whether the respective variables of two systems show the same trend of changes or not, regardless of the magnitudes of the respective variables.

\begin{figure}
	\centering
	\includegraphics[width=\columnwidth]{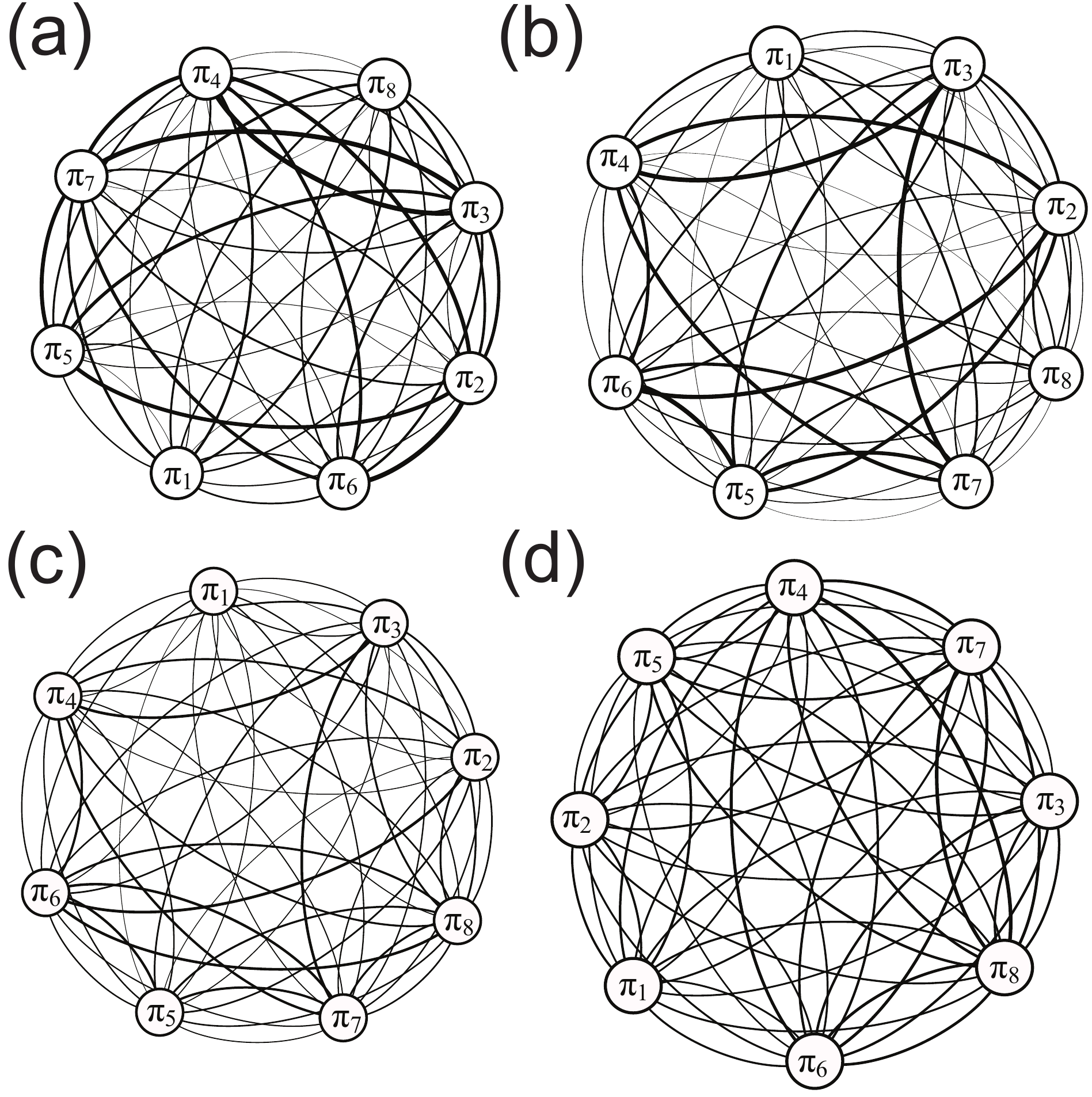}
	\caption{\small{Ordinal pattern transition network for (a) a single stochastic process. (b) a COPT for two coupled stochastic processes, (c) an alternative version of COPT, and (d) a JOPT. For better visualization, the arrows of links are suppressed.} \label{fig:stochasticCOPT}}
\end{figure}

A direct quantitative comparison between a COPT and a JOPT seems not possible since we have different definitions for patterns. Anyway, we present qualitative similarities when we show a COPT and a JOPT for two interacting stochastic processes in Fig. \ref{fig:stochasticCOPT}. In particular, we consider two stationary processes $X_t$ and $Y_t$, each of which admits an autoregressive representation
\begin{align}
X_t &= \sum_{j=1}^{p} a_{2j} X_{t-j} + \sum_{j=1}^{p} b_{2j} Y_{t-j} + \eta_{1t}, \\
Y_t &= \sum_{j=1}^{p} c_{2j} X_{t-j} + \sum_{j=1}^{p} d_{2j} Y_{t-j} + \eta_{2t},
\end{align}
where the noise terms $\eta_{1t}, \eta_{2t}$ are uncorrelated, $p$ is the order, $a_{2j}, b_{2j}, c_{2j}$ and $d_{2j}$ are model coefficients. The terms of $b_{2j}$ and $c_{2j}$ reflect the interacting strength between $X_t$ and $Y_t$ and therefore if $X_t$ and $Y_t$ are independent, $b_{2j}$ and $c_{2j}$ are uniformly zero. Here we consider a simple case of $p = 2$, $a_{2j} = 0.5$, $b_{2j}=0.1$,  $c_{2j} = 0.5$, and $d_{2j}=0$, when there is only unidirectional interactions from $Y_t$ to $X_t$.

In order to keep the same number of ordinal patterns as defined in Tabs. \ref{tab:3DCOPT} and \ref{tab:3DJOPT}, we reconstruct a three dimensional phase space for $X_t$ (respectively $Y_t$) using time delay embedding techniques ($\tau = 1$). Figure \ref{fig:stochasticCOPT}a shows the ordinal pattern transition network for a single stochastic process where we find rather random transition patterns in the resulting network. These random transitions between patterns have been observed in the COPT and JOPT (Figs. \ref{fig:stochasticCOPT}b, c and d). When increasing the interacting strength term $b_{2j}$, one would expect some reduced level of random pattern transitions in the $X_t$ process because of the unidirectional interactions from the $Y_t$ process. The dependence on the interaction strength requires further investigations, which is beyond the topic of the current work.

\subsection{Permutation entropy}
In the next step, we quantify the heterogeneous properties of the resulting networks. For a deterministic system, the frequencies of ordinal patterns are generally different from each other and the existence of forbidden patterns is simply a special case in this regard. Traditionally, permutation entropy $\mathcal{H}$ is introduced to characterize the inhomogeneous appearance of ordinal patterns as following
\begin{equation}
	\mathcal{H}_O = - \sum_{i=1}^{2^n} p(\pi_i) \log_2 p(\pi_i),
\end{equation}
where the sum runs over all $D = 2^{n}$ permutations and $n$ is the dimension of one system and $p(\pi_i)$ is the probability of order pattern $\pi_i$. We use the observation frequency $\mathcal{F}(\pi_i)$ to estimate $p(\pi_i)$ and furthermore we use $\log_2$ and  hence the units of $\mathcal{H}_O$ are bits. For a $n$-dimensional independent identical distributed stochastic process, one obtains the largest entropy $\mathcal{H}_O = n$ since each of $D = 2^{n}$ ordinal patterns is expected to have the same frequency.

The computation of $\mathcal{H}_O$ characterizes the different frequencies of order patterns and it has been well demonstrated that the transition behavior between ordinal patterns is not fully captured by $\mathcal{H}_O$ \cite{zhangSciRep2017}. To this end, we first indicate each directed link representing the order pattern transitions in the resulting network by its transition frequency $w_{ij} = p({\pi_i \to \pi_j})$, following the time iterations of the series. In order to emphasize the importance of non-self transitions between ordinal patterns, self-loops have been removed as suggested \cite{zhangSciRep2017}. Finally, we obtain a weighted directed network characterized by a weighted adjacency matrix $W = \{ w_{ij} \}, i, j \in [1, 2^{n}]$. The matrix $W$ fulfils the normalization $\sum_{i,j}^{2^n} w_{ij} = 1$. Here, based on $W$, the regularity of the order pattern transition properties is quantified by the Shannon entropy $\mathcal{H}_T$, which is
\begin{equation} \label{eq:HtS}
\mathcal{H}_T = - \sum_{i,j=1}^{2^{n}} w_{ij} \log_2 w_{ij},
\end{equation}
where the sum runs over all possible $2^{2n}$ transitions. In a full analogy to $\mathcal{H}_O$, for a $n$-dimensional independent identical distributed stochastic process, one obtains the largest entropy $\mathcal{H}_T = 2 n$.

Both measures have been demonstrated to show the capabilities to capture the different bifurcation transition scenarios. In the examples of this work, we show that $\mathcal{H}_O$ and $\mathcal{H}_T$ are sensitive to PS.

\section{Detecting transitions to PS} \label{secSync}
\subsection{Preliminaries on PS}
In this section, traditional measures characterizing PS are briefly reviewed and more historical details can be found in \cite{PecoraChaos2015,BoccalettiPR2002,Pikovsky_Kurths_synchr}. PS is characterized by the phase locking $|m \phi_1 - n \phi_2 | < C$, where $m$, $n$ and $C$ are constants and $\phi_1(t) $ and $\phi_2(t) $ are phases of the two oscillators. In the case of two coupled R\"ossler oscillators, we easily compute phases by $\phi_1(t) = \arctan {y_1(t)}/{x_1(t)}$ and $\phi_2(t) = \arctan{y_2(t)}/{x_2(t)}$ since the systems are in phase coherent regimes. In addition, $m$ and $n$ are often chosen as $1$ because $1:1$ phase synchronization is more often observed. Equivalently, the phase locking is characterized as the average frequency mismatch drops to zero, namely, $\Delta \Omega = 0$, where $\Delta \Omega = \Omega_1 - \Omega_2$ and $\Omega_{1,2} = \frac{1}{2\pi} \left< \frac{d \phi_{1,2}(t)}{d t} \right>$, where $\left< \cdot \right>$ is an average over time $T$. PS has been related to the spectrum of Lyapunov exponents \cite{RosenblumPRL1997} and it is characterized by the transition of the Lyapunov exponent from zero to negative values \cite{Pikovsky_Kurths_synchr}. 

\subsection{Unidirectional coupling schemes}
In the coupled R\"ossler systems (Eqs. \ref{eq_roessler}), PS can be achieved by applying the unidirectional coupling scheme $\kappa_1=0$ and relative small frequency mismatch $\Delta$ between $\omega_1$ and $\omega_2$ \cite{zhengPRE2000}. More specifically, we choose $\Delta = 0.02$ and $(\omega_1$, $\omega_2) = (0.98, 1.02)$. Note that in this model, generalized synchronization is obtained for relative large values of $\Delta = 0.2$ which leads to $(\omega_1, \omega_2) = (0.8, 1.2)$. The interrelationship between PS and generalized synchronization has been systematically investigated in \cite{zhengPRE2000}. 

\begin{figure}
	\centering
	\includegraphics[width=\columnwidth]{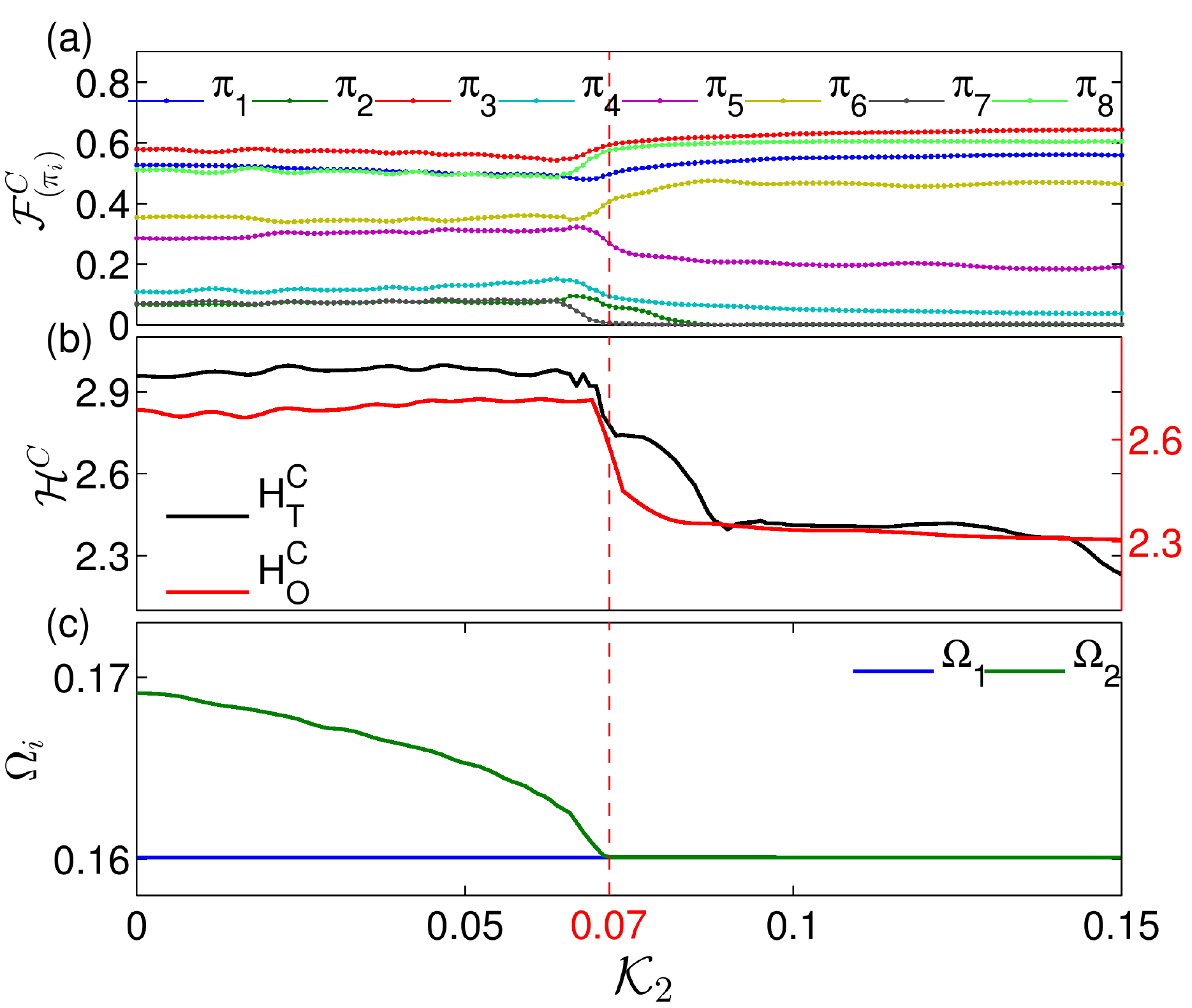}
	\caption{\small{Network measures of COPTs versus the coupling strength $\kappa_2$ for two unidirectionally coupled R\"ossler systems ($\kappa_1 = 0$, $\Delta = 0.02$, namely, $\omega_1 = 0.98$ and $\omega_2 = 1.02$ in Eqs. \ref{eq_roessler}. (a) frequencies of ordinal patterns $\mathcal{F}(\pi_i)$ while increasing the coupling strength $\kappa_2$. (b) $\mathcal{H}_O$ and $\mathcal{H}_T$. (c) The average rotation frequencies $\Omega_1$ and $\Omega_2$. In all cases, vertical dashed highlight the critical synchronization transition thresholds $\kappa_2 = 0.07$ \cite{zhengPRE2000}. Superscripts $C$ denote COPTs. } \label{fig:twoRosslerW1COPTJOPT}}
\end{figure}
\begin{figure*}
	\centering
	\includegraphics[width=\textwidth]{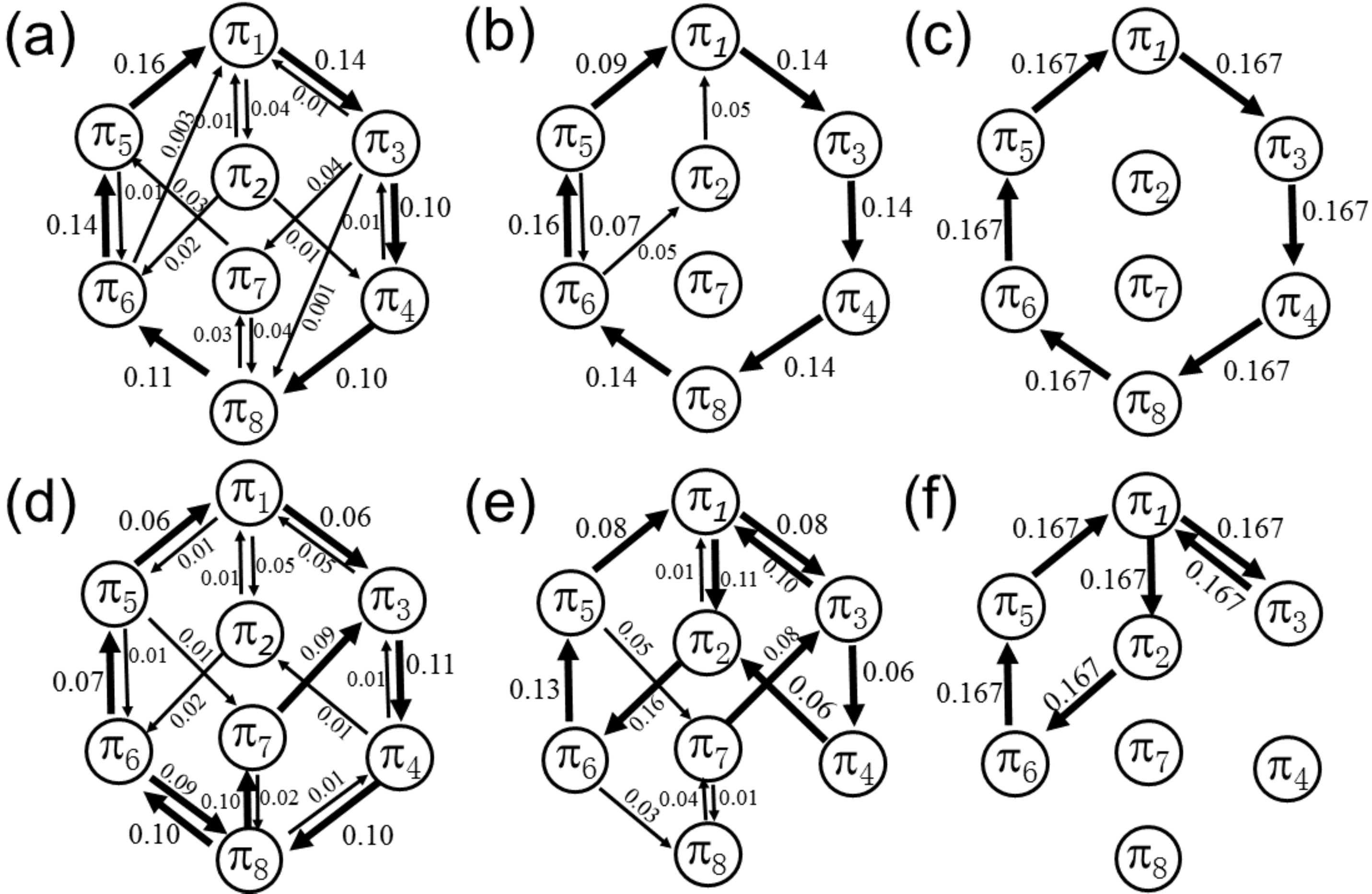}
	\caption{\small{Network illustrations for COPTs (a-c) and JOPTs (e-f). (a, d) $\kappa_2 = 0.04$, (b, e) $\kappa_2 = 0.08$, and (c, f) $\kappa_2 = 0.12$. Directions of pattern transitions and the corresponding frequencies are indicated by arrows and the values on links. } \label{fig:networkCOPJOP}}
\end{figure*}

We first shown in Fig. \ref{fig:twoRosslerW1COPTJOPT} that the variations of network measures of COPTs depending on the coupling parameters $\kappa_2$ ($\kappa_1 = 0$). When increasing the coupling $\kappa_2$, frequencies of ordinal patterns $\mathcal{F}(\pi_i)$ experience relative large variations (Figs. \ref{fig:twoRosslerW1COPTJOPT}(a)). In consequence, both $\mathcal{H}_{T}$ and $\mathcal{H}_{O}$ show fast decays to small values when the coupling threshold $\kappa_2$ passing the transition values (as highlighted by vertical dashed lines in Fig. \ref{fig:twoRosslerW1COPTJOPT}(b)). The transition point to PS has been validated by the average rotation frequencies of $\Omega_1$ and $\Omega_2$ since they are locked to the same value at $\kappa_2 = 0.07$. We choose three representative coupling values $\kappa_2=0.04, 0.08$, and $0.12$ to show the structural variations of the corresponding COPTs in Fig. \ref{fig:networkCOPJOP}(a-c). In the non-synchrony regime ($\kappa_2 = 0.04$), the transitions between ordinal patterns and their frequencies are rather random (Fig. \ref{fig:networkCOPJOP}(a)). As the coupling increases to just above the critical value $\kappa_2 = 0.08$, a dominant (more deterministic) transition route emerges (Fig. \ref{fig:networkCOPJOP}(b)). When PS is achieved for large $\kappa_2 = 0.12$, more missing patterns have been observed (Fig. \ref{fig:networkCOPJOP}(c)). We note that the different pattern definitions are used between the ordinal transition network for a single chaotic R\"ossler system as shown in Fig. \ref{fig:rosCOPT}(a) and the COPT for two coupled systems (Fig. \ref{fig:networkCOPJOP}(c)). The unidirectional coupling leads to the entrainment of the response to the drive system, which yields the same ordinal partitions for phase space as the drive. Due to the continuity of phase space trajectory, we observe a similar deterministic transition route between phase space partitions as for a single system. 

\begin{figure}
	\centering
	\includegraphics[width=\columnwidth]{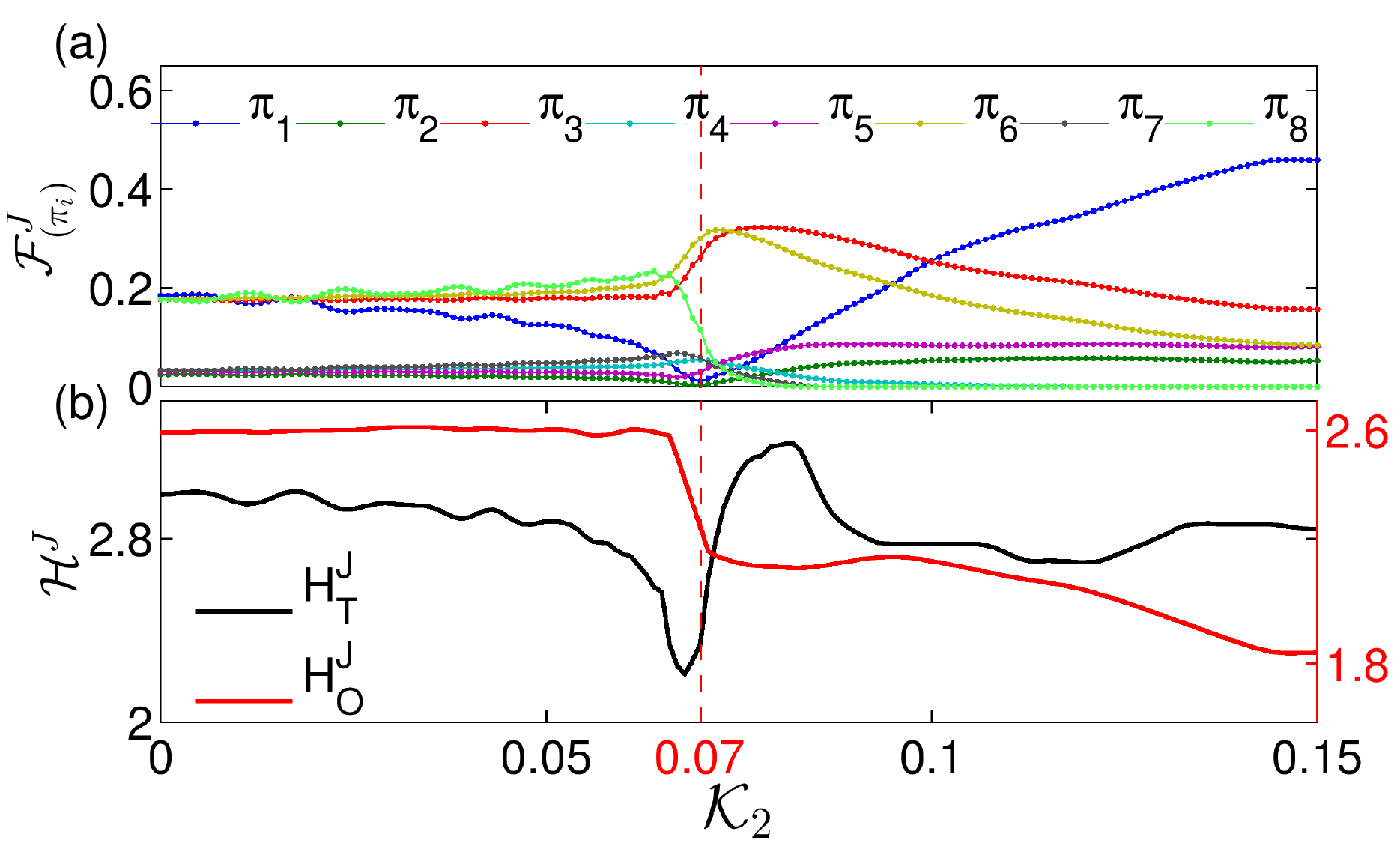}
	\caption{\small{Network measures of JOPTs versus the coupling strength $\kappa_2$ for two unidirectionally coupled R\"ossler systems ($\kappa_1 = 0$). (a) Frequencies of ordinal patterns while increasing the coupling strength. (b) $\mathcal{H}_O$ and $\mathcal{H}_T$. Superscripts $J$ denoted JOPTs. } \label{fig:twoRosslerW2COPTJOPT}}
\end{figure}

Concerning network measures from JOPTs, we have obtained rather similar results as shown in Fig. \ref{fig:twoRosslerW2COPTJOPT}. Three representative JOPTs on the route to PS are illustrated in Fig. \ref{fig:networkCOPJOP}(d-f). As the coupling strength increases, the random transitions between patterns become more deterministic yielding more missing patterns. 

Based on the results of Figs. \ref{fig:twoRosslerW1COPTJOPT}, \ref{fig:twoRosslerW2COPTJOPT} and \ref{fig:networkCOPJOP}, we conclude that $\mathcal{H}_O$ and $\mathcal{H}_T$ can effectively capture the transitions to PS. However, the detailed variations of particular patterns remain unclear during the transitions to PS. In other words, we have to look into the details of the appearance or disappearance of each ordinal patterns (as shown in Figs. \ref{fig:twoRosslerW1COPTJOPT}(a) and \ref{fig:twoRosslerW2COPTJOPT}(a)) because some of the patterns become suppressed, which provide further evidence in characterizing the synchronization dynamics. 

It is easy to understand the occurrences of missing patterns on the transition routes to synchronization. As the coupling strength is increased during the synchronization transitions, the high dimensional coupled systems are constrained to a lower dimensional synchronization manifold which means that the determinism of the system is increased. In consequence, both the frequencies of order patterns and the transitions between different patterns become more inhomogeneous. Therefore, we compute the missing probability (frequency) of each pattern while varying the coupling strengths. 

\subsection{Bidirectional coupling schemes}
In this section, we implement bidirectional coupling scheme when choosing $\kappa_1 = \kappa_2 = \kappa$. In addition, we show the transition to synchronization in a two dimensional parameter space of coupling strength $\kappa \in [0.0, 1.2]$ and natural frequency mismatch $\Delta \in [-0.04, 0.04]$. We choose this particular range of parameters because there are different routes of chaos-chaos, chaos-period-chaos transitions to synchronizations \cite{mamenPLA2004}. This space $(\Delta, \kappa)$ is further divided into $800 \times 1200$ grid points by equal step size. For each parameter combination, we integrate the ODEs with the same strategy as described in Sec. \ref{secRungeK} to obtain time series for the coupled system. Then, we construct both COPTs and JOPTs. 

\begin{figure*}
	\centering
	\includegraphics[width=\textwidth]{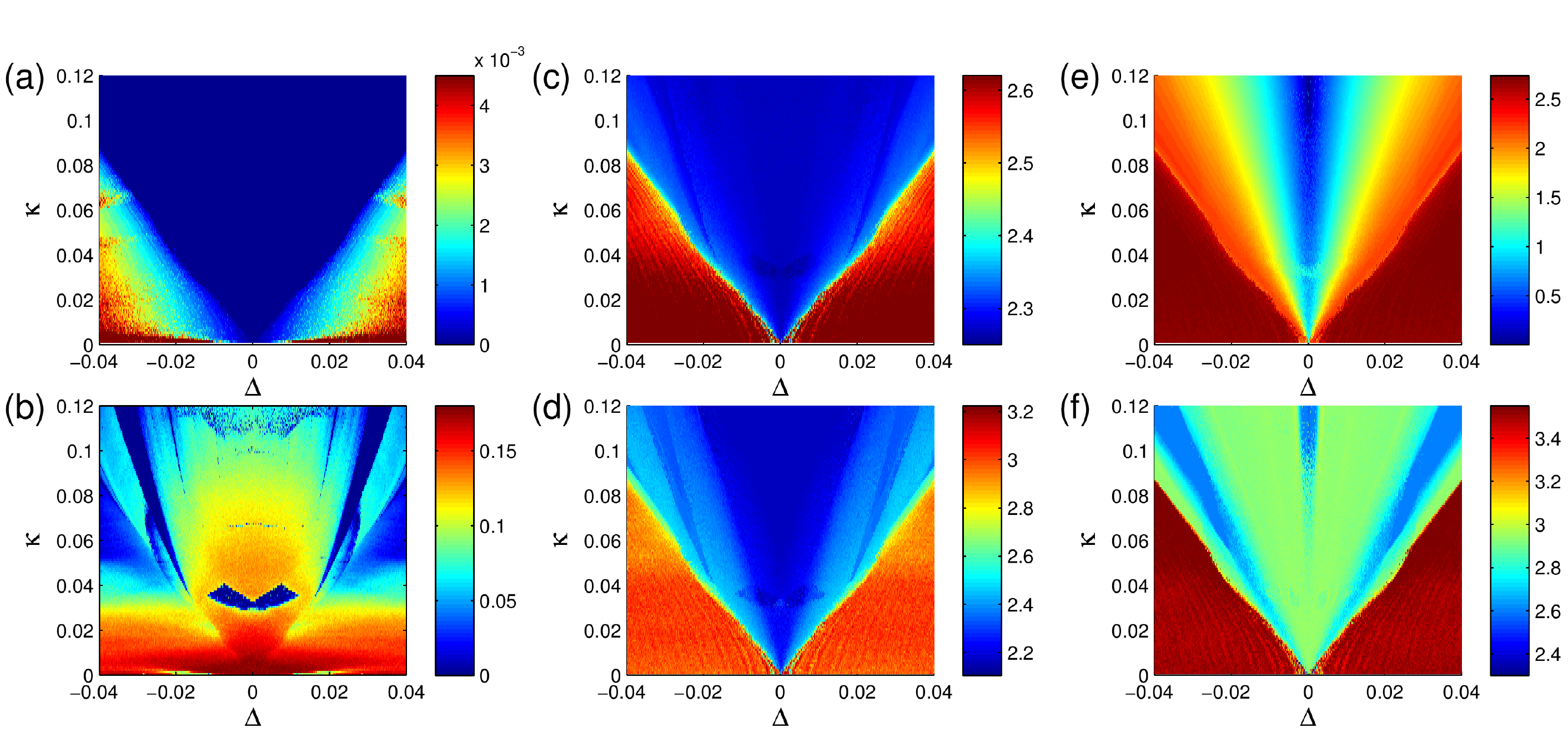}
\caption{\small{(color online) Two parameter space (frequency mismatches $\Delta$ and coupling strength $\kappa_1 = \kappa_2 = \kappa$), which is color coded by the network statistics obtained from COPTs and JOPTs. (a) the difference of the mean frequencies $\Delta \Omega = \Omega_1 - \Omega_2$, (b) the sum of non-negative Lyapunov exponents of the coupled system. COPTs: (c) $\mathcal{H}_O$, (d) $\mathcal{H}_T$, and JOPTs: (e) $\mathcal{H}_O$, (f) $\mathcal{H}_{T}$. } \label{fig:twoparaCOPTOne}}
\end{figure*}

First, we show the average frequency mismatches $\Delta \Omega$ in the parameter space $(\Delta, \kappa)$ and PS is characterized by the well-known Arnold tongue \cite{Pikovsky_Kurths_synchr}, as shown in Fig. \ref{fig:twoparaCOPTOne}(a). PS are observed inside this Arnold tongue, while no PS outside this region. We further validate these results by computing the Lyapunov spectrum of the whole system based on Eqs. \ref{eq_roessler}  (Fig. \ref{fig:twoparaCOPTOne}(b)). We note that it may not be possible to distinguish the tip of the Arnold tongue only by considering the sum of positive Lyapunov exponents because of the non-hyperbolic property of the coupled $6$-dimensional system. One interesting structure inside the Arnold tongue is the periodic region close to $0.03 < \kappa < 0.04$ (like two eyes), which has been reported in \cite{mamenPLA2004}. It is often claimed that all transitions between different types of synchronization are related to the changes in the Lyapunov spectrum \cite{RosenblumPRL1997}. In particular, for low values of the coupling strength, one has the following configuration for the coupled $6$-dimension system: $\{ \lambda_1 > 0, \lambda_2 > 0, \lambda_3 \sim 0, \lambda_4 \sim 0, \lambda_5 < 0, \lambda_6 < 0 \}$. Increasing the coupling strength, PS is achieved when $\lambda_4$ becomes negative. If the coupling strength is further increased, generalized synchronization is obtained as $\lambda_2 \sim 0$ and $\lambda_3 < 0$. However, for intermediate coupling strengths and relative large frequency mismatches, the spectrum of Lyapunov exponents has limited power in explaining the weak correlations outside the Arnold tongue, although the phases are not locked \cite{mamenPLA2004}. 

Figures \ref{fig:twoparaCOPTOne}(c-f) show the parameter space color coded by the entropy values $\mathcal{H}_O$ and $\mathcal{H}_T$, which are calculated from the respective COPTs and JOPTs. The outer boarders of the Arnold tongue have been successfully captured by $\mathcal{H}_{O}$ and $\mathcal{H}_{T}$, comparing to the average frequency mismatch plot (Fig. \ref{fig:twoparaCOPTOne}(a)). More importantly, all these entropy values are able to capture the tip of the Arnold tongue, however, the spectrum of Lyapunov exponents does not (Fig. \ref{fig:twoparaCOPTOne}(b)). Therefore, Figs. \ref{fig:twoparaCOPTOne}(c-f) provide important complementary information to the sum of the positive Lyapunov exponents. 

Inside the Arnold tongue, however, these entropy values drop to small values, presenting more intricate gradient structures. Anyway, entropy values do not show convincingly the periodic region close to $0.03 < \kappa < 0.04$ except for the case of $\mathcal{H}_{T}$ computed from COPTs. This is because $\mathcal{H}_O$ and $\mathcal{H}_T$ are averaged over all patterns. 

In order to illustrate more clearly the gradient structures inside and on the edge of the Arnold tongue, we plot the missing (failure) probabilities of individual ordinal patterns in the space $(\Delta, \kappa)$. For one combination of parameter pair $(\Delta, \kappa)$, we run $n=200$ simulations of random initial conditions. The COPT and JOPT have been reconstructed for each realization of $N=800000$ time points and the frequency plots of ordinal patterns have been obtained. Furthermore,  this ensemble of independent realizations results in a failure with probability $p_{\pi_i} = 1$ if pattern $\pi_i$ is missing and a success with probability $p_{\pi_i} = 0$ if $\pi_i$ is observed, respectively.  

\begin{figure*}
	\centering
	\includegraphics[width=\textwidth]{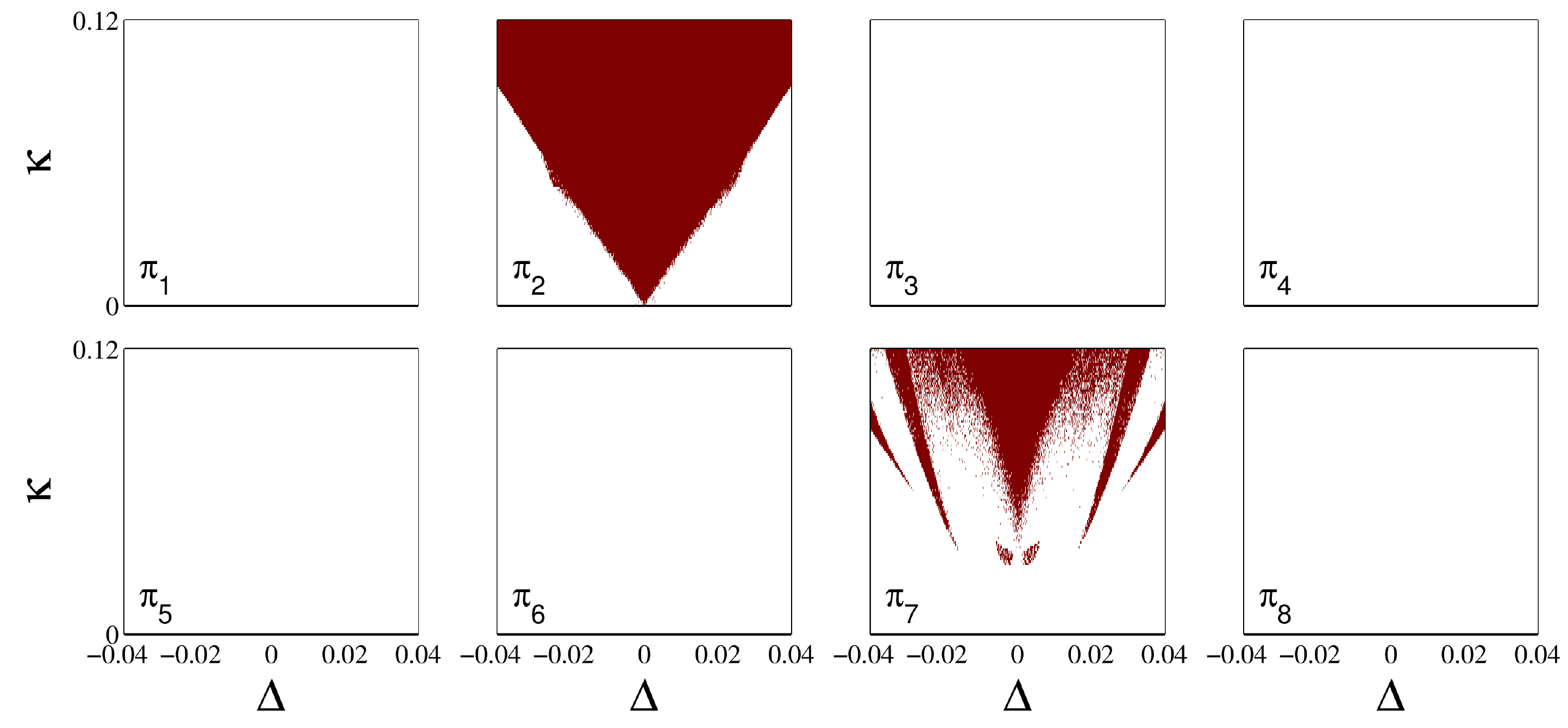}
\caption{\small{(color online) Two parameter space $(\Delta, \kappa)$ highlighted by the missing (failure) probability of individual pattern obtained from COPTs. Note that $\pi_1, \pi_3, \pi_4, \pi_5, \pi_6$ and $\pi_8$ are observed for the entire range of parameters (white areas). Red regions correspond to the missing probabilities of patterns, which show that $\pi_2$ captures mainly the outer boarders of the Arnold tongue and $\pi_7$ captures the inside structures including two eye-like periodic regions. } \label{fig:twoparaCOPTTwoC}}
\end{figure*}

\begin{figure*}
	\centering
	\includegraphics[width=\textwidth]{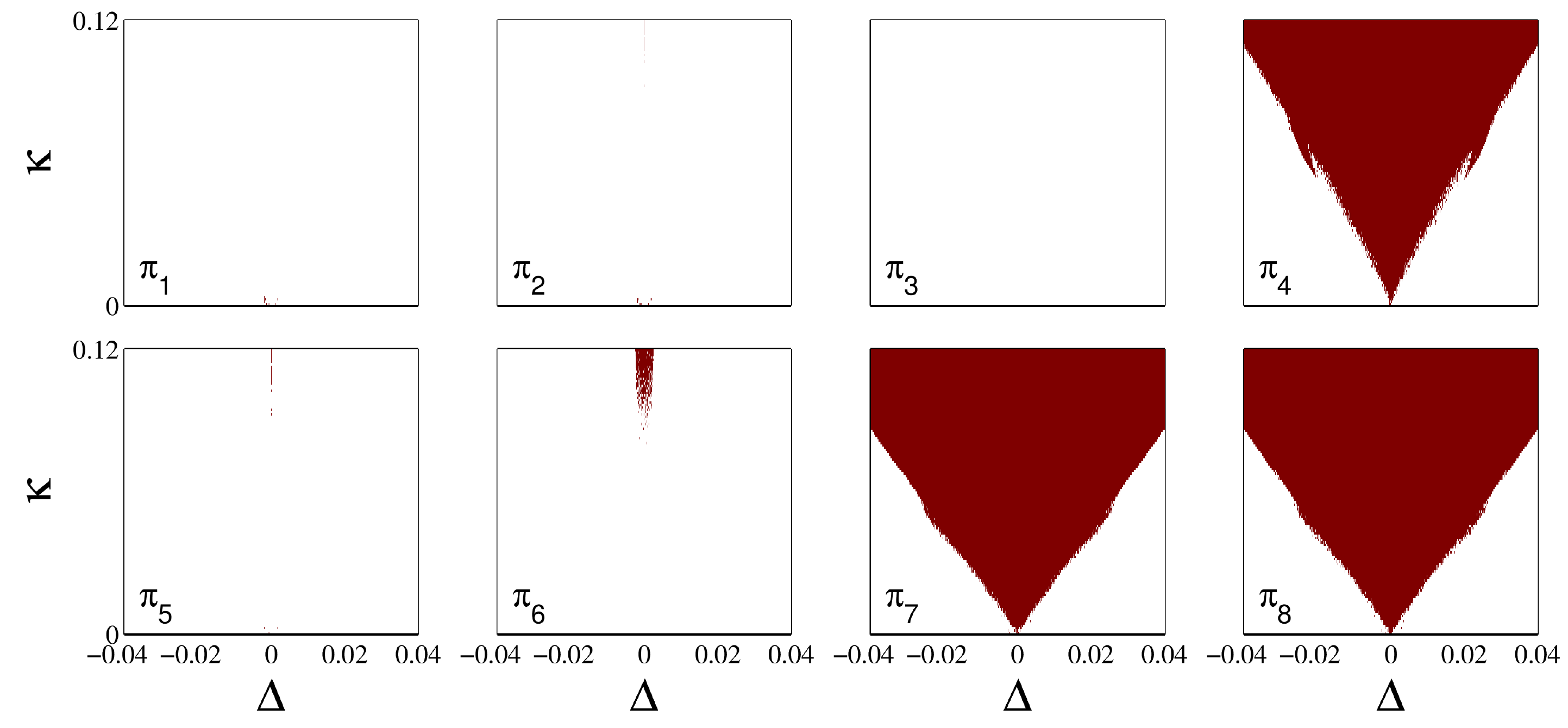}
\caption{\small{(color online) Two parameter space $(\Delta, \kappa)$ highlighted by the missing (failure) probability of individual pattern obtained from JOPTs. Note that $\pi_1, \pi_2, \pi_3$ and $\pi_5$ are observed for the majority area of the parameter space (white areas). Red regions correspond to the missing probabilities of patterns showing that $\pi_7$ and $\pi_8$ capture the outer boarders, while $\pi_4$ and $\pi_6$ affect the inside structures of the Arnold tongue. } \label{fig:twoparaCOPTTwoJ}}
\end{figure*}

In the case of COPTs, Fig. \ref{fig:twoparaCOPTTwoC} shows that $\pi_1, \pi_3, \pi_4, \pi_5, \pi_6$ and $\pi_8$ are observed patterns for the entire range of parameters we considered. In addition, we find that pattern $\pi_2$ is completely not observed in the Arnold tongue and the disappearance of $\pi_2$ determines sharply the outer boarder. In addition, pattern $\pi_7$ plays a key role in forming the periodic eye structures inside the Arnold tongue. In the case of JOPTs, Fig. \ref{fig:twoparaCOPTTwoJ} shows that patterns $\pi_1, \pi_2, \pi_3$ and $\pi_5$ do appear in the majority area of the parameter space. Again, the disappearance of patterns $\pi_7$ and $\pi_8$ determine sharply the outer boarders of the Arnold tongue, while the inside structures are determined by $\pi_4$ and $\pi_6$. 

\section{Conclusions} \label{secCons}
In summary, we propose to construct cross and joint ordinal pattern transition network from multivariate time series, which has been particularly applied to analyze synchronization transitions. Note that a COPT and JOPT are two slightly different ways to construct networks from time series, providing complementary information. The ordinal patterns of a COPT are defined by considering the signs of the difference of $\Delta \vec{x}_1 - \Delta \vec{x}_2$ between two subsystems. In contrast, the ordinal patterns of a JOPT  are defined by the signs of the product of $\Delta \vec{x}_1 \cdot \Delta \vec{x}_2$. It is certain that the amplitudes of oscillations of different variables influence directly the definition of a COPT. However amplitudes become not important for a JOPT because only the signs of the product are considered. In addition, it is straightforward to generalize the ideas of JOPTs from two to three (or even $n$) coupled subsystems with an extended number of pattern definitions. We plan to further generalize the ideas of JOPTs to multilayer or multiplex networks for time series analysis. However, it remains to be a big challenge for constructing a COPT for three coupled subsystems. 

Based on the cross and joint ordinal pattern transition networks, we propose two entropy measures to characterize the resulting networks. Namely, $\mathcal{H}_O$ is calculated from the frequencies of each patterns and $\mathcal{H}_T$ is obtained from the transition frequencies between any pair of patterns. Our results show that both $\mathcal{H}_O$ and $\mathcal{H}_T$ track successfully the critical coupling threshold to phase synchronization. The applications of our method to generalization synchronization analysis is a more challenging task and will be a subject for future work \cite{PecoraChaos2015}. 

In the two parameter space of $(\Delta, \kappa)$, both entropy measures capture the tip of the Arnold tongue successfully, providing complementary information to the traditional measure of Lyapunov exponents. In order to show the intricate structures inside the Arnold tongue, our results suggest that we should study the missing probability of each pattern separately, instead of relying on the global measures of $\mathcal{H}_O$ and $\mathcal{H}_T$. This is because individual pattern shows different sensitivity to dynamic transitions. In particular, we have observed particular missing patterns which correspond to the outer boarders and the inner structures in the parameter space. 

In addition to synchronization analysis based on real time series, it remains to be an interesting topic to identify the driver-response relationship, especially to identify indirect from direct coupling directions \cite{rosenblum_pre2001,RomanoPRE2007,Nawrath_prl_2010,zouIJBC2011}. From the viewpoint of ordinal pattern perspective, it is possible to combine ordinal recurrence plots \cite{grothPRE2005}  and cross and joint ordinal partition transition network approaches to tackle this problem. More importantly, we plan to address the statistical significance of the coupling directions for real time series data. 

\section{Acknowledgement}
This work is in part financially sponsored by Natural Science Foundation of Shanghai (Grants No. 17ZR1444800 and 18ZR1411800).


\end{document}